\documentclass[sn-mathphys,Numbered]{sn-jnl}

\usepackage{graphicx}%
\usepackage{multirow}%
\usepackage{amsmath,amssymb,amsfonts}%
\usepackage{amsthm}%
\usepackage{mathrsfs}%
\usepackage[title]{appendix}%
\usepackage{xcolor}%
\usepackage{textcomp}%
\usepackage{manyfoot}%
\usepackage{booktabs}%
\usepackage{algorithm}%
\usepackage{algorithmicx}%
\usepackage{algpseudocode}%
\usepackage{listings}%
\usepackage{subcaption}%
\usepackage{graphicx}%

\usepackage{float}

\begin{document}

\title[Article Title]{An Overview of Quantum Software Engineering in Latin America}


\author*[1]{\fnm{Álvaro M.} \sur{Aparicio-Morales}}\email{amapamor@unex.es}
\equalcont{These authors contributed equally to this work.}

\author[1]{\fnm{Enrique} \sur{Moguel}}\email{enrique@unex.es}
\equalcont{These authors contributed equally to this work.}

\author[2]{\fnm{Luis Mariano} \sur{Bibbo}}\email{lmbibbo@lifia.info.unlp.edu.ar}
\equalcont{These authors contributed equally to this work.}

\author[2,3]{\fnm{Alejandro} \sur{Fernandez}}\email{casco@lifia.info.unlp.edu.ar}
\equalcont{These authors contributed equally to this work.}

\author[1]{\fnm{Jose} \sur{Garcia-Alonso}}\email{jgaralo@unex.es}
\equalcont{These authors contributed equally to this work.}

\author[1]{\fnm{Juan M.} \sur{Murillo}}\email{juanmamu@unex.es}
\equalcont{These authors contributed equally to this work.}

\affil*[1]{\orgdiv{Department of Computer and Telematic System Engineering}, \orgname{University of Extremadura}, \orgaddress{\street{Avenida de la Universidad, s/n}, \city{Cáceres}, \postcode{10004}, \state{Extremadura}, \country{Spain}}}

\affil[2]{\orgdiv{LIFIA, Facultad de Informática}, \orgname{Universidad Nacional de La Plata}, \orgaddress{\street{Calles 50 y 120}, \city{La Plata}, \postcode{1900}, \state{Province of Buenos Aires}, \country{Argentina}}}

\affil[3]{\orgname{Comisión de Investigaciones Científicas de la Provincia de Buenos Aires}, \orgaddress{\street{Calle 526 e/ 10 y 11}, \city{La Plata}, \postcode{1900}, \state{Province of Buenos Aires}, \country{Argentina}}}



\abstract{Quantum computing represents a revolutionary computational paradigm with the potential to address challenges beyond classical computers' capabilities. The development of robust quantum software is indispensable to unlock the full potential of quantum computing. Like classical software, quantum software is expected to be complex and extensive, needing the establishment of a specialized field known as Quantum Software Engineering. Recognizing the regional focus on Latin America within this special issue, we have boarded on an in-depth inquiry encompassing a systematic mapping study of existing literature and a comprehensive survey of experts in the field. This rigorous research effort aims to illuminate the current landscape of Quantum Software Engineering initiatives undertaken by universities, research institutes, and companies across Latin America. This exhaustive study aims to provide information on the progress, challenges, and opportunities in Quantum Software Engineering in the Latin American context. By promoting a more in-depth understanding of cutting-edge developments in this burgeoning field, our research aims to serve as a potential stimulus to initiate pioneering initiatives and encourage collaborative efforts among Latin American researchers.}




\maketitle

\section{Introduction}
\label{sec1}

Quantum computing is postulated as a new computing paradigm capable of addressing several challenges that classic computers cannot \cite{Piattini2020}, such as factorizing prime numbers \cite{salm2020nisq}, simulation of chemical reactions (quantum simulation), or resolving meteorological equations \cite{Gaitan2020}, among others. This is possible due to its quantum nature, superposition, and entanglement properties. These characteristics give quantum computers a huge computation capacity that fits with several applications, such as cryptography, medicine, and energy. Nevertheless, to develop those applications in a large-scale and industrial way, quantum software is needed \cite{Piattini2020}. 

As it happens in nowadays' software solutions, quantum software is expected to become large and complex to create high-quality solutions or applications. However, the necessary elements for developing it are still in a very early stage \cite{Moguel2022}. Given this context, Quantum Software Engineering (QSE) is presented as a necessary field to achieve that goal as proposed in the Talavera Manifesto \cite{Piattini2020} in which a synopsis outlining certain principles and commitments regarding Quantum Software Engineering and Programming is presented. 

This new area, QSE, focuses on bringing the benefits of actual software engineering (SE) to quantum software development. In this way, efforts in researching, training, and organizing activities and/or events that promote the exchange of ideas and knowledge around QSE are necessary. 

Leveraging the special issue focused on Latin America, it is important to know the development made by universities, research institutes, and companies to highlight these advancements in Quantum Software Engineering. This can be a starting point for pioneering initiatives and collaborations between Latin American countries to improve and develop this new area of quantum software engineering. An initial example of such initiatives can be found at the RIPAISC network\footnote{\url{https://www.ripaisc.net} ``Red Iberoamericana para el Avance de la Ingeniería de Software Cuántico''}.

In this regard, it is fundamental to establish cooperation between those stakeholders. Hence, it is necessary to identify who is working around QSE and their research domain inside this new field. Therefore, our research tries to provide a first approximation toward understanding the current landscape of Quantum Software Engineering research in Latin America. Additionally, we seek to offer an initial assessment of participants' perspectives on the prevalent issues and challenges in QSE, to promote research and development in this field.

For this purpose, we conducted a systematic mapping study of Quantum Software Engineering, identifying research works from Latin American universities or research centers to explore their research interests in Quantum Software Engineering. In addition, we surveyed experts in the field to capture information about affiliations, involvement with QSE, funding and resources, collaboration, participation in events and publications, opportunities, and challenges.

To detail this information the rest of the paper is structured as follows. Section 2 briefly introduces Software Engineering, its procedures and practices, and their adaptation to Quantum Software development. Section 3 presents the methodology and results of the systematic literature mapping. Section 4 describes the survey we conducted to collect information on the current state of QSE in Latin America and the results derived from this survey. Finally, Section 5 details the conclusions of this work.


\section{Background}
\label{sec2}

Software has evolved from simple tasks executed on rudimentary computers to large and complex programs that operate in a distributed manner. Achieving this required establishing guidelines and techniques to create large-scale and industrial software that meets various demands for seamless execution. Over time, these guidelines and techniques became a discipline called Software Engineering \cite{IEEE1990}. 

As the Talavera Manifesto \cite{Piattini2020} states, there is a rapidly increasing awareness of the need for quantum computing applications and quantum software. Therefore, it is time to adapt, modify, or even create new methodologies and processes within the field of Software Engineering.

In this section, Software Engineering and its procedures and practices will be introduced and explained as well as its adaptation to Quantum Software development.
  

\subsection{Software Engineering}
\label{subsec1}

Various authors have formulated formal definitions for the Software Engineering (SE) discipline, such as the definition of \textit{Andriole} and \textit{Freeman} \cite{Andriole1993}, \textit{Sommerville} \cite{Sommerville2011} or \textit{Boehm} \cite{Boehm1976} among others. In this paper, the considered as \emph{standard} definition of SE provided by IEEE \cite{IEEE1990} is taken, which is as follows:

\emph{The application of a systematic, disciplined, quantifiable approach to the development, operation, and maintenance of software; that is, the application of engineering to software. (2) The study of approaches as in (1).}

As can be inferred from its definition, the software engineering process encompasses all aspects of software, from development to maintenance. This implies a collection of methods and rules applied throughout the software's lifetime. This iterative process, occurring over the lifetime of the software, is also known as the Software Lifecycle. 

The Software Lifecycle sometimes called the Software Development Lifecycle (SDLC), is a collection of stages or phases the software goes through during its lifetime. These phases cover from its conception, via the definition of requirements to decommissioning.

Thanks to the processes carried out at each of these stages in software development, it becomes possible to create large and complex software that is robust, fault-tolerant, and of high quality \cite{Basiri2016}.

\begin{figure}[h]
    \centering
    \includegraphics[width=0.6\textwidth]{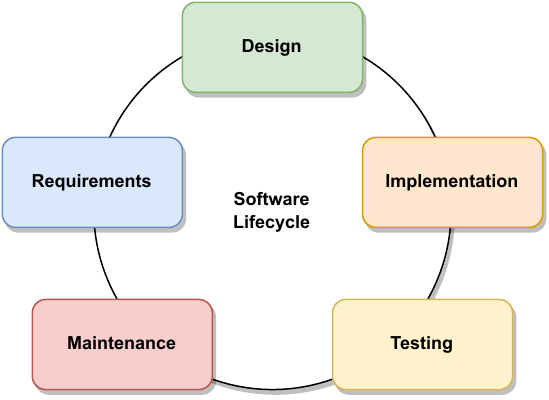}
    \caption{Software Development Lifecycle (SDLC)}
    \label{fig:Software_Development_Lifecycle}
\end{figure}

A diagram of the software development life cycle stages is shown in Figure \ref{fig:Software_Development_Lifecycle}. Each stage represents an area of study and application in software engineering.

Given the actual state of quantum computers, a way of using them is through their integration with classical computers, forming what is termed hybrid systems (classical-quantum) \cite{Rojo2021}. In a hybrid system, each computer will possess its own software. Consequently, both kinds of software, classical and quantum, will coexist. In this context, it is logical to presume that similarly to how software engineering is applied to classical software, applying software engineering processes to quantum software would be necessary. However, because of the difference between both computing paradigms, an adaptation or, in some cases, a transformation of the stages and practices in the Classical Software Development Lifecycle to this new computing paradigm and programming is necessary. 

Before defining the concept of Quantum Software Engineering, some basic ideas of differences between classical and quantum software will be presented for a better understanding.

To this end, we need to understand what is meant by Software. Software can be defined as the functional logic of a computer system. Until now, ``logic'' has primarily been employed to denote computational principles rooted in Boolean algebra (0-1). However, this new computing paradigm presents a new form of logic: Quantum Logic. Quantum Logic differs from the one formulated by \textit{George Boole}. Still, the concept of computational logic should not be modified since both logic (quantum and classical) have the same purpose: to provide behavior to the computer system (either quantum or classical).

Taking this into account, the concept of software is not inherent to the computing model that the computer system possesses, therefore, if (in short) the definition of software engineering is the application of engineering processes to software development, this definition can be maintained in the field of quantum computing, bearing in mind that software can now be both, quantum and classical.

Given this context, the explanation of the concept of Quantum Software Engineering is the same as for Classical Software Engineering by IEEE. Just keep in mind that the concept of software currently associated with Classical Computing must also be associated with Quantum Computing.

The following sections provide a more detailed overview of Quantum Software Engineering, outlining the activities involved in each phase.

\subsubsection{Software Requirements}
\label{subsubsec1}

In Software Engineering the stage of software requirements can be defined as the initial phase where the needs, functionalities, and constraints of a new or modified software system are identified, documented, and analyzed in detail.

In this step, software engineers collect the expected behavior of the functionalities within our software system, encompassing the specific actions and operations it should perform (functional requirements). Additionally, software engineers consider the system's operation circumstances, including performance, security, usability, or other quality attributes (non-functional requirements). These requirements not only define what the system should do but also specify how it should behave under certain conditions to ensure proper functionality and performance. As a result, a document in which each (functional and non-functional) requirement is collected and defined is obtained. 

Identifying and describing a functional requirement in Quantum Software will be done similarly to today's. However, there are minor differences concerning non-functional requirements. Given the current restrictions in quantum computers, there are several aspects to take into account to executing a software program in a quantum computer, such as the noise, the availability of a computer, the number of qubits and shots, or the cost of execution, which are inherent features of this new paradigm.

This means that the quantum computer in which the software will be executed must support the determined number of shots, qubits, and a maximum noise level according to the established non-functional requirements.

\subsubsection{Software Design}
\label{subsubsec2}

The software design stage consists of modeling an abstract representation of the software and generating an architecture, structure, and components. The model must consider all the requirements of the previous stage and be a guide to the people who implement, test, and maintain the software.

At the end of this phase, an architectural model and software specifications are defined. To obtain the architectural model, a modeling language is necessary. In modern Software Engineering, the Unified Modelling Language (UML) is considered a standard in software design \cite{Zhao2020}. Applying this \emph{standard} in the design of quantum software seems plausible. Some works are starting to appear, such as the one of \textit{Pérez-Castillo et al.} \cite{PerezDelgado2020} proposing an extension of UML for two kinds of diagrams (class and sequence). On the other hand, the work of \textit{Ali et al.} \cite{Shaukat2020} introduced an initial conceptual model of quantum software from the standpoint of model-based engineering. 

As mentioned before, the software specification also needs to be generated at the end of this step. A specification formally expresses a computer program in a language software engineers understand. This language is known as a specification language. In software engineering, it is utilized for modeling the behavior of computer programs. In the same way that extensions of the language used for modeling are necessary for the architectural model, an extension will also be necessary to apply specification language to quantum software. In this line, we can highlight the article by \textit{Cartiere} \cite{Cartiere2022}, which introduces a formal specification that can represent the basic notations of quantum computing.

\subsubsection{Software Implementation}
\label{subsubsec3}

In Software Engineering, the implementation phase centers on creating the code according to the requirements, specifications, and model obtained in the previous steps. It involves translating the design blueprints into functional software components and validating their adherence to the established guidelines. 

In classical software, the code generation is made by using a specific programming language, frameworks, and tools that allow and speed up solution development. Moreover, one of the characteristics of classical software is that it is not a monolithic artifact. Instead, it adopts a modular and distributed structure in which different components work in a coordinated and autonomous way to create a comprehensive solution for a particular problem \cite{AparicioMorales2023}.

In modern systems, such a distributed architecture consists of small software modules called microservices. This way of development allows the use of different methods, programming languages, and tools to implement each system's components. The distributed architecture enables the creation of a scalable and efficient solution that can also be containerized using specialized tools and allows it to be deployed on any computer.

Achieving this advance in computing programming has been possible due to a set of factors such as the democratization of classical computers, years of development to create libraries and packages to compose complex software as well as software tools to make developing computer programs without forgetting the application of engineering to the entire software development process. 

Nowadays, quantum software programming is mainly based on using quantum gates to implement existing quantum algorithms. In the last few years, new advanced algorithms have been developed due to the increasing number of qubits available in the new generation of quantum computers. Nevertheless, some problems may delay the advancement of quantum software development. Among them, the following can be highlighted:

\begin{itemize}
    \item Noise: Errors and inaccuracies arising during quantum computing executions can propagate and affect the reliability of the computations performed by the quantum software. Having these problems and errors in the experimental stage of developing software is acceptable, and it is necessary to face them and find a solution. However, encountering software errors during the production stage could lead to severe issues, especially if this software is utilized in critical areas such as electrical plants, bank transactions, or high-precision surgeries.
    
    \item Accessibility: Nowadays, access to quantum computers is mostly through cloud providers. They offer users a specified availability window when the quantum computer is online. Additionally, the quantum algorithm uploaded to the cloud provider is placed in a queue, awaiting execution. Usually, this causes long waiting times to collect the results and analyze them to check if they are working adequately or if the algorithm needs improvement.
    
    \item Economic cost: Because of the novelty of quantum technology, executing algorithms in a Cloud environment becomes costly. Consequently, this can be an obstacle for individuals, research groups, or companies with few financial resources allocated for their quantum software execution to be forced to limit attempts in running their quantum programs on actual quantum computers. 
\end{itemize}

Additionally, there are various providers offering access to different quantum computers. This results in many development libraries, each specific to its provider and often using different programming languages. Consequently, the lack of standardization may impede progress in quantum software development.   

However, despite these disadvantages, significant efforts are being made to address these current issues to facilitate the implementation and usage of quantum software for developing quantum-classical systems.

There are several lines of development for the implementation of quantum software, including the work by \textit{Rojo et al.} \cite{Rojo2021}, \textit{Moguel et al.} \cite{Moguel2022}, and \textit{Valencia et al.} \cite{Valencia2022} focus on the development of a concept known as hybrid microservices. These involve encapsulating a quantum algorithm within a microservice framework based on classical technologies.

Another line of work consists of developing tools that make the process of programming a quantum algorithm easier. For instance, \textit{Romero-Álvarez et al.} \cite{Romero2022} demonstrate a method to define quantum services and automatically generate the corresponding source code by extending the Open API Specification. Another noteworthy contribution is the work conducted by \textit{Garcia-Alonso et al.} \cite{GarciaAlonso2022}, where the authors introduce an adaptation of the API Gateway pattern, considering the unique aspect that quantum services cannot be deployed in the same manner as traditional services. Instead, their Quantum API Gateway suggests the optimal quantum computer to execute a particular quantum service during runtime.

Therefore, it is crucial to have tools operating at higher abstraction levels to streamline the construction of intricate hybrid systems and widen quantum programming accessibility. Furthermore, addressing the essential skills, capabilities, and toolsets needed by future quantum software programmers is imperative.

\subsubsection{Software Testing}
\label{subsubsec4}

In the software testing phase, the software developed during the implementation phase is thoroughly examined to identify defects and errors and to ensure that it meets the specified requirements and quality standards. Software testing aims to validate and verify that the software functions correctly, performs as expected, and delivers the intended outcomes.

To test a classical computer program, it undergoes a series of tests that examine each function within the program. Due to the deterministic nature of classical programming, verifying program results is a straightforward task. By knowing the expected result value and comparing it with the real result value obtained after the execution of a function, it can be determined whether the function is operating correctly.  To conduct a test, it is necessary to specify the inputs, the execution conditions, the testing procedure, and the expected results. After executing all the tests to the software, if none fails, it indicates that everything is working as expected. Consequently, it can be concluded that the computer program has passed the test suite and is ready for the next step of the software development lifecycle.

With quantum computers, the main phase of the test when verifying if a function is working as expected is quite different because of the probabilistic nature of the results. This characteristic calls for a different approach to testing, using methods and practices that consider the uncertainties in this new computing paradigm. During the verification process, evaluating statistical probabilities and distributions of outcomes becomes necessary. 

Another aspect to consider is that in classical programming, there exists the possibility of adding a breakpoint in the software. A breakpoint in computer programming is ``\emph{a point in a computer program at which execution can be suspended to permit manual or automated monitoring of program performance or results}" \cite{IEEE1990}. When the execution of a program is halted at a breakpoint, the testing software engineer can inspect the actual state of the computer program.  Furthermore, this tool allows them to proceed with the execution of the software, either line by line, function by function, or from one breakpoint to another. Such functionality greatly aids testers in debugging by providing insights into why the computer program might malfunction. Given the quantum mechanics nature of this new computing paradigm, conducting a measurement significantly influences the observation \cite{Hidary2019}. This implies that placing breakpoints in the computer program to know the state of the system and perform debugging tasks is impossible and it will be necessary to look for new alternative methods.

The previous aspects pose considerable challenges to modifying or creating new testing methodologies for the quantum software. 

In quantum software testing, one of the relevant works is that of \textit{Miranskyy et al.} \cite{Miranskyy2019}. They addressed several challenges linked to white-box and black-box testing, along with verifying and validating quantum programs. Another interesting work is the one proposed by \textit{de la Barrera et al.} \cite{delaBarrera2022} in which strategies of classical software testing are discussed for being applied to quantum software testing.

\subsubsection{Software Maintenance}
\label{subsubsec5}

Maintenance is the last step of the Software Development Lifecycle. During this stage, continuous software evolution occurs through bug fixing, making enhancements, updates, and addressing user feedback to ensure the smooth operation of the software during its lifetime.

Today's software is not only made up of the computer program that is developed but also makes use of different libraries and packages external to it, which also undergo continuous updates and improvements in terms of algorithmic efficiency, security, and even new functionalities that can affect the software correct performance. Therefore, the computer program must be maintained to function correctly with the libraries and packages that make it up.

Over time, the same thing will happen to quantum software as it happens to software today. More and more libraries and packages will appear with functionalities that will make the development of quantum software easier, and they will receive updates that may affect the quantum software's correct performance.

Nowadays, low-level updates to the software of quantum computers can interfere with the \emph{correct} execution of our algorithm. Moreover, due to the variety of quantum computers implementing different hardware technologies, the result in the execution of our program may differ from one provider to another.  Thus, in these early stages of quantum computing and quantum software development, performing proper maintenance of our quantum programs is necessary. 

Another aspect of this stage of the software development life cycle is the continuous improvement of the processes performed by our computer program, either through new functionalities or through improvements to existing functionalities, the latter is known as the re-engineering process.

Due to the coexistence of both types of software, certain functionalities currently performed in classical software may eventually be migrated to quantum software using the re-engineering process due to reductions in execution times. Also, new quantum functionality will be integrated into existing classical software. This is not a simple task, and it would be necessary to develop standards, models, and systematic transformations to preserve business rules and simplify migration across various classical and quantum environments \cite{Piattini2021}. 

There has been limited effort in applying reverse engineering techniques to quantum software within maintenance stages. However, \textit{Pérez-Castillo et al.} \cite{PerezCastillo2020} addresses this issue by proposing a software modernization approach. This approach aims to restructure classical systems alongside existing or new quantum algorithms, aiming to create target systems that combine classical and quantum information systems.

\section{Systematic Mapping Study}
\label{sec3}

The main objective of this study is to collect information of interest on the current State of the Art of Quantum Software Engineering in Latin America. For this purpose, a systematic mapping study is carried out where, in addition to reaching the proposed objectives, it is sought to classify recent research works in Quantum Software Engineering by Latin American researchers.

A systematic mapping study is a type of secondary study (study based on the analysis of previous research). Its objective is to determine the scope of research conducted on a specific research topic and to classify knowledge, as opposed to a systematic review that seeks to answer a specific research question \cite{Kitchenham2011,Petersen2008}.

We will use this technique to search the defined scope, classify the most relevant research, and conduct a thematic analysis to offer a visual map of knowledge of the main research in Quantum Software Engineering in Latin America.

For this purpose, we follow the methodology proposed by \textit{Petersen et al.} \cite{Petersen2008} in which they suggest a procedure consisting of 5 stages: \ref{sec3.1} Define research questions; \ref{sec3.2} Perform a literature search; \ref{sec3.3} Select studies; \ref{sec3.4} Classify papers; and \ref{sec3.5} Extract and perform data aggregation.

Each defined step was carried out by the researchers signing this document. To improve readability and facilitate understanding of the results obtained, we decided to focus only on explaining the classification of the papers (stage \ref{sec3.4}) and the analysis (stage \ref{sec3.5}) in the following sections.

\subsection{Research Questions}
\label{sec3.1}

Although the general objective of this study could be summarized as understanding what techniques, methodologies, and technologies exist in Quantum Software Engineering in Latin America, this objective is divided into three specific research questions to obtain more detailed knowledge and a global vision of the subject. The research questions to be answered in this study are the following: \\

\textbf{RQ1. What techniques and/or technologies exist in the field of Quantum Software Engineering in Latin America?} \\

\textbf{RQ2. Is there evidence of interest in Software Engineering applied to Quantum Computing in Latin American countries?} \\

\textbf{RQ3. What are the research areas and problems in which Quantum Software Engineering is being applied to in Latin America?}

\subsection{Data Sources and Search Strategy}
\label{sec3.2}

The search strategy was developed taking into account the specific terminology related to Software Engineering for Quantum Computing. To perform the search, terms such as software, program, computation, circuit, or engineering and their main synonyms were considered. The search string was as follows:

\begin{lstlisting}
    quantum 
    AND 
    (software OR program* OR comput* OR circuit OR enginee*)
\end{lstlisting}

The search was conducted in the Scopus\footnote{\url{https://www.scopus.com}} and IEEE\footnote{\url{https://ieeexplore.ieee.org/Xplore/home.jsp}} databases. These libraries have a wide coverage of publications in computer science (among others) and index several publication catalogs (including other libraries such as ACM, Springer, Elsevier, etc.).

\subsection{Study Selection}
\label{sec3.3}

Four simple criteria were defined for study selection:

\begin{itemize}
    \item Inclusion. All papers dealing with any aspect related to Software Engineering or nearby, such as tools, techniques, etc., were considered.
    
    \item Exclusion. We decided to eliminate all those papers that use this new computing paradigm but do not deal with any aspect related to engineering. Additionally, and for logical reasons, duplicate articles were discarded.

    \item Constraint. Only the papers in which at least one of the authors is from a Latin American country (Argentina, Brazil, Chile, Colombia, Costa Rica, Ecuador, El Salvador, Guadeloupe, Honduras, Mexico, Paraguay, Peru, Puerto Rico, Uruguay or Venezuela) were taken into account.

    \item Timing. The concept of Quantum Software Engineering was coined by \textit{Stepney et al.} in 2004 \cite{Stepney2004}. However, it was not until the concept of Quantum Supremacy was defined by \textit{John Preskill} \cite{Preskill2012} that QSE took on special relevance. For this reason, we have considered restricting the search from 2012 onwards. In addition, the interest of our proposal is to know the most active researchers in recent years, which is why we have restricted the search to the last 11 years.
\end{itemize}

\subsection{Paper Classification}
\label{sec3.4}

Performing the search in the indicated scientific databases and applying the inclusion and exclusion criteria already detailed, the study's results revealed 1,316 works. The complete list of works analyzed in this paper can be found as supplementary material on the Zenodo platform\footnote{\url{https://doi.org/10.5281/zenodo.10718726} Supplementary material\label{first_footnote}}.

To facilitate the understanding of the analysis of the articles and due to the high number of proposals, the classification was carried out in three phases: 1) only the title, abstract, and keywords of each paper were analyzed according to the inclusion and exclusion criteria (detailed in the previous section); 2) the same elements of each paper were evaluated to classify them according to the Software Engineering topic; and 3) the full text of the papers were analyzed to refine the selected papers further to answer the defined research questions comprehensively.

After the classification phase, the total number of papers classified as software engineering papers is 91 out of 1,316 (approximately 6.9\%). This is a low number, but understandable because most of the papers from the search deal with this new computing paradigm applied to different contexts of physics, chemistry, cybersecurity, finance, etc., and do not refer to any aspect related to Software Engineering. Details of this selection of papers can be found as supplementary material on the Zenodo platform\footref{first_footnote}.

\begin{figure}[H]
    \centering
    \includegraphics[width=0.8\textwidth]{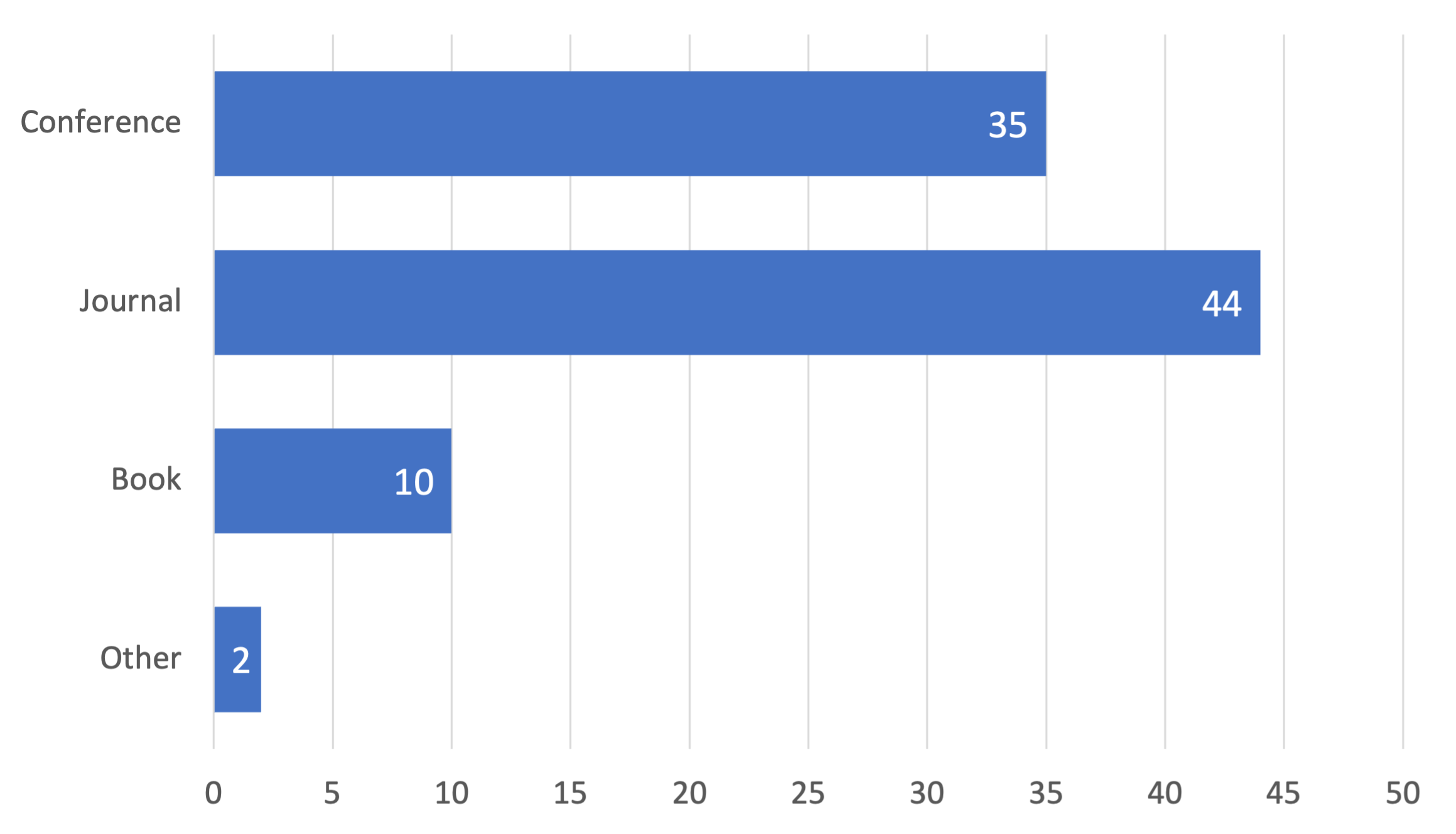}
    \caption{Venue types}
    \label{fig:Type}
\end{figure}

These papers were published in different venues. Figure \ref{fig:Type} presents a distribution of articles in terms of publishing venues. As it can be appreciated, 79 out of 91 Software Engineering papers were published in Journals and Conferences, standing out the publications of Journals with a total of 44 out of 91 papers, approximately 48.3\%. Only 13.1\% represent publications in books and other media.

\begin{figure}[h]
    \centering
    \includegraphics[width=0.8\textwidth]{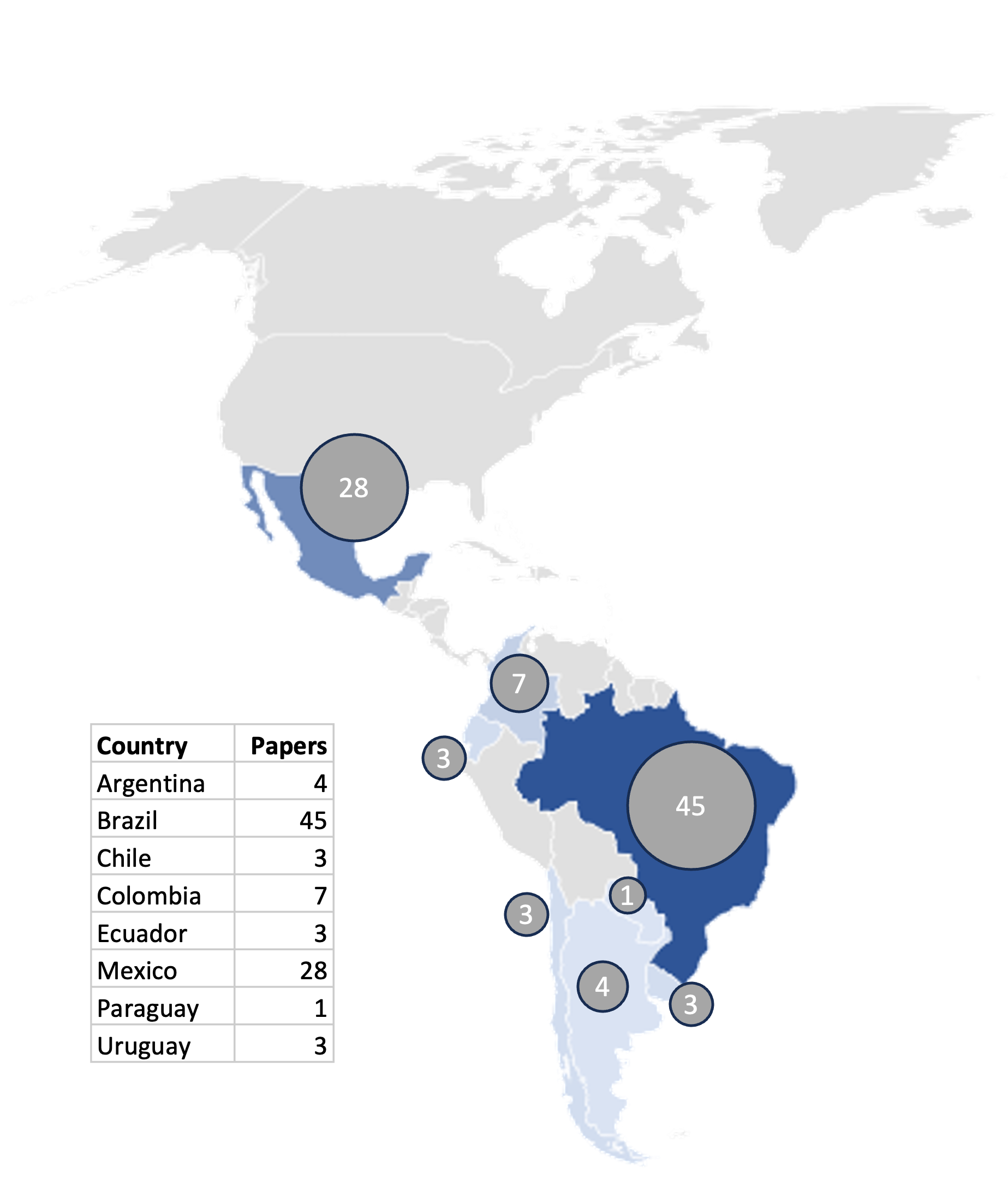}
    \caption{Bubble map by country}
    \label{fig:BlobMap}
\end{figure}

On the other hand, in Figure \ref{fig:BlobMap}, a geographical distribution of the Latin American authors of the published papers is shown. A total of 8 Latin American countries are currently working on projects related to Software Engineering applied to Quantum Computing. Among the countries, two of them represent approximately 80.2\% of the published papers. These countries are Brazil, with a total of 45 papers, and Mexico, with 28 papers. 

It is worth mentioning that in Brazil, three authors can be highlighted because they are the most prolific ones. They are \textit{Adenilton J. da Silva}, \textit{Teresa B. Ludermir}, and \textit{Marley Vellasco}, with 3 selected works each. \textit{Da Silva} and \textit{Ludemir} (2 selected works) usually work together because they have collaborations and their area of study in Software Engineering applied to Quantum Computing is the development of Neural Networks and development methodologies \cite{DePaulaNeto2020,Araujo2023}.

In addition, in Mexico, the most prolific ones are \textit{César N. Plata}, \textit{Raymundo Marcial Romero}, and \textit{Antonio Hernández Servín}, with 3 selected works each. These three researchers work together to develop extensions for the quantum programming language QML (Qt Modelling Language) \cite{PlataCesar2020,PlataCesar2019,PlataCesar2018}. Also, we would like to highlight that in Argentina, from the 4 selected papers, 3 of them are from the researcher \textit{Alejandro Diaz-Caro}, who works on formal aspects of quantum programming \cite{DiazCaro2018,DiazCaro2017,DiazCaro2014}.

In addition, we would like to highlight that there are collaborations between researchers from different Latin American countries. Of the classified works, there are 3 papers in which collaborations exist between researchers from Argentina and Brazil \cite{Sousa2020}, Brazil and Chile \cite{Dinani2023}, and between Chile and Ecuador \cite{Cardenas2019}.

\subsection{Data Extraction and Synthesis}
\label{sec3.5}

For a systematic mapping study, \textit{Kitchenham et al.} \cite{Kitchenham2011} and \textit{Petersen et al.} \cite{Petersen2008} defined guidelines suggesting that other parts of the analyzed articles should only be evaluated in cases where they are not well structured or are imprecise. For this research, we decided to fully analyze each manuscript to answer all the questions raised better. Some important proposals could be left out by analyzing only titles, abstracts, and keywords.

\begin{figure}[h!]
    \centering
    \includegraphics[width=0.8\textwidth]{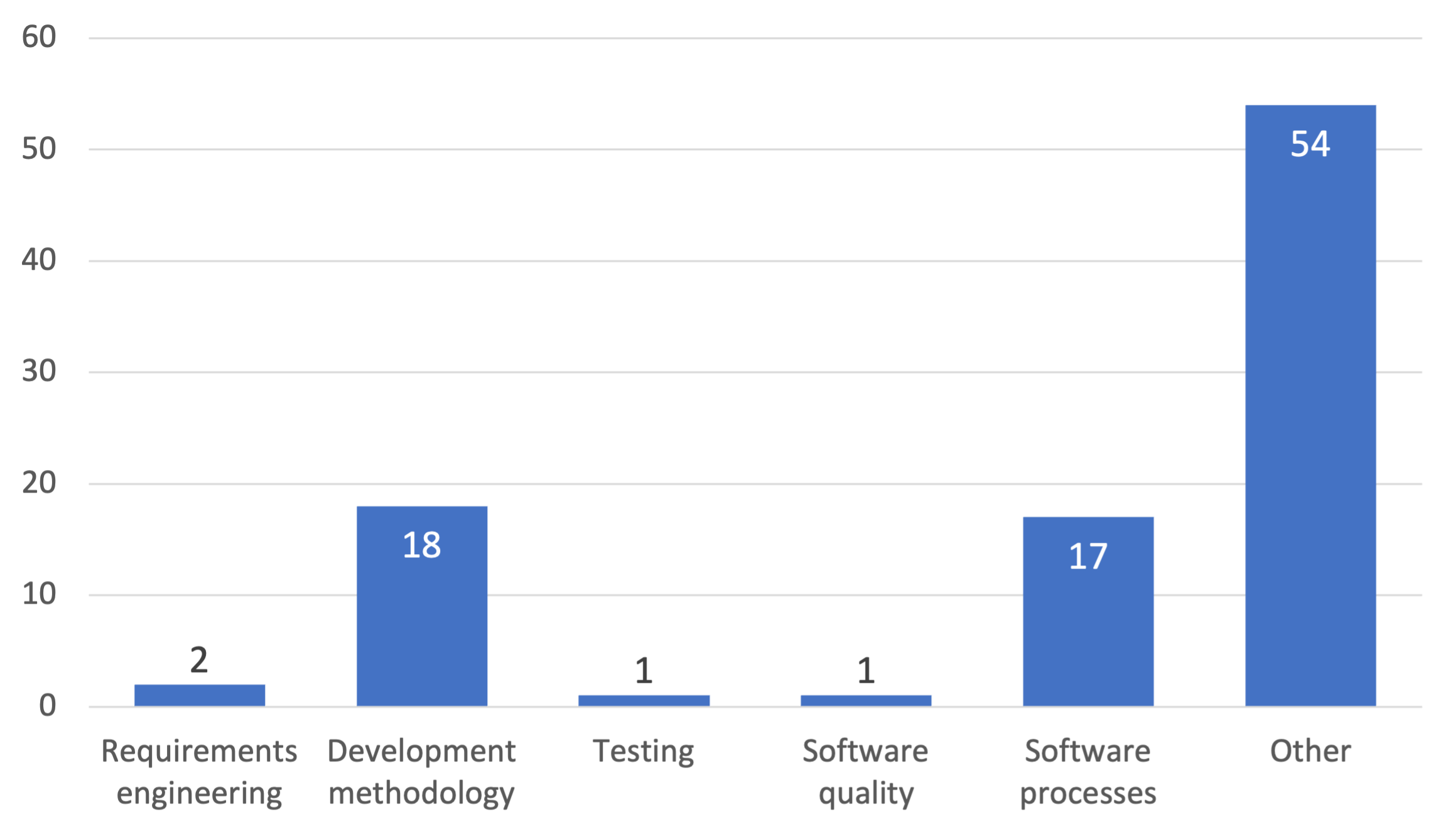}
    \caption{Categories of Software Engineering}
    \label{fig:Category}
\end{figure}

Classifying the papers according to the different fields of Software Engineering, the result is shown in Figure \ref{fig:Category}. It can be seen that most of the papers that exist so far are in the field of Development Methodologies (18 papers) and Software Processes (17 papers). Also, we have included a category of other works because most of them are works that make use of methodologies, techniques, or tools in conjunction with quantum computing, and not being software engineering its purpose.

As shown in Figure \ref{fig:Category}, two papers address \textbf{Requirements Engineering}. In the first one, \textit{Saraiva et al.} \cite{Saraiva2021} analyzes the non-functional requirements of quantum programs based on experiences in quantum software development for real quantum hardware. Also, the authors analyze hardware-related constraints and derive a set of generic non-functional requirements for this type of program, identifying five non-functional requirements related to performance efficiency and reliability that should be considered when implementing a quantum program for a quantum device. The second of the papers, \textit{Diaz et al.} \cite{Diaz2022}, shows a study of different design techniques and implementations to simulate quantum computing with the support of classical computing, highlighting quantum resource consumption considerations.

Regarding \textbf{Development Methodologies}, we have found 18 papers. Of all of them, we highlight the work of \textit{Feitosa et al.} \cite{Feitosa2019,Feitosa2016} in which they propose a monadic semantics for FJQuantum, an object-oriented language based on Featherweight Java, created to reason and develop programs that handle quantum data and quantum operations. The authors also provide an interpreter for this semantics and examples of quantum programs that run with it. On the other hand, we would also like to highlight the work of \textit{Gejea et al.} \cite{Adugna2023} in which they present a high-performance simulation methodology of the Grover method that combines the advantages of the Grover technique with the parallel of cloud computing to improve multicore load balancing, memory utilization, and simulation efficiency.

On the other hand, in specific articles dedicated to \textbf{Testing}, we have only come across the work of \textit{Diaz et al.} \cite{Diaz2022} who perform some tests on the tools that they implement. Given its nature and the topics it deals with, this work was the only one we have classified into 2 categories (Requirements Engineering and Testing).

Similarly, in the field of \textbf{Software Quality}, we have only found the work of \textit{Oliveira et al.} \cite{Oliveira2022} in which they propose a framework to identify the sensitivity of quantum circuits to radiation-induced faults and the probability that a fault in a qubit propagates to the output. The researchers build on recent studies and experiments on real quantum machines, modeling transient faults in a qubit as a phase shift with a parameterized magnitude.

The Software Engineering category with the highest number of articles was \textbf{Software Processes} with 17 papers. Of these works, we would like to highlight the work of \textit{Dias et al.} \cite{Dias2013} in which they present a process based on QIEAs (quantum-inspired evolutionary algorithms) that make use of the evolution of machine code programs. Multilevel quantum systems inspire this process. Its operation is based on quantum individuals, representing a superposition of all programs (solutions) in the search space. On the other hand, we would also like to highlight the work of \textit{Silva et al.} \cite{Silva2021} in which they propose a software process for mapping the traveling salesman problem (TSP) based on pseudo-Boolean constraints to the D-Wave Systems. The authors formulate the problem as a set of constraints in propositional logic. Finally, we would also like to highlight the work of \textit{Ballinas et al.} \cite{Ballinas2022} in which they propose a process based on HQGA (Hybrid Quantum Genetic Algorithm) for the 0-1 Knapsack Problem (KP). The authors rely on the principles of quantum computation (superposition and entanglement of states) and use the Qiskit simulator.

In this study, we decided to go beyond the merely strict in terms of Software Engineering, and we wanted to evaluate other aspects related to methodologies, techniques, or tools in Quantum Computing that are applied in diverse areas. For them, we have decided to incorporate an additional category called \textbf{Others}. This category mainly includes works that make use of Neural Networks for different fields of application, for example, \cite{Leal2021} or \cite{DePaula2020}; make use of Machine Learning techniques for different projects as \cite{Payares2023}, \cite{Hossein2023} or \cite{Trevor2022}; or algorithmic or cybersecurity issues as \cite{Yadav2023} or \cite{Singamaneni2022}; among other types of works.

\section{Surveying LATAM Researchers' Involvement in QSE}
\label{sec4}

The main objective of this survey is to collect information about the present state of Quantum Software Engineering in Latin America. To gather this information, we created a survey using the Google Forms platform and sent it to a group currently working in research institutions in Latin America. The survey and the anonymized survey responses can be found as supplementary material on the Zenodo platform\footnote{\url{https://doi.org/10.5281/zenodo.10718726} Supplementary material}.

This survey was sent to the 18 members of the RIPAISC network\footnote{\url{https://www.ripaisc.net} ``Red Iberoamericana para el Avance de la Ingeniería de Software Cuántico''} and to 113 researchers who were referred to us by the Network's researchers. This survey was also sent to the 39 Latin American researchers of the papers extracted from the Literature Mapping Study detailed in Section \ref{sec3}. A massive mailing of our survey was sent to 152 researchers. No question was defined as mandatory to make it easier for respondents to answer.


We obtained 34 responses from the Latin American scientific community. Below, we will detail the most important aspects extracted from the responses received.


\subsection{Participation}

\begin{figure*}[h!]
    \centering
    \begin{subfigure}[b]{0.48\textwidth} 
        \centering    \includegraphics[width=\textwidth]{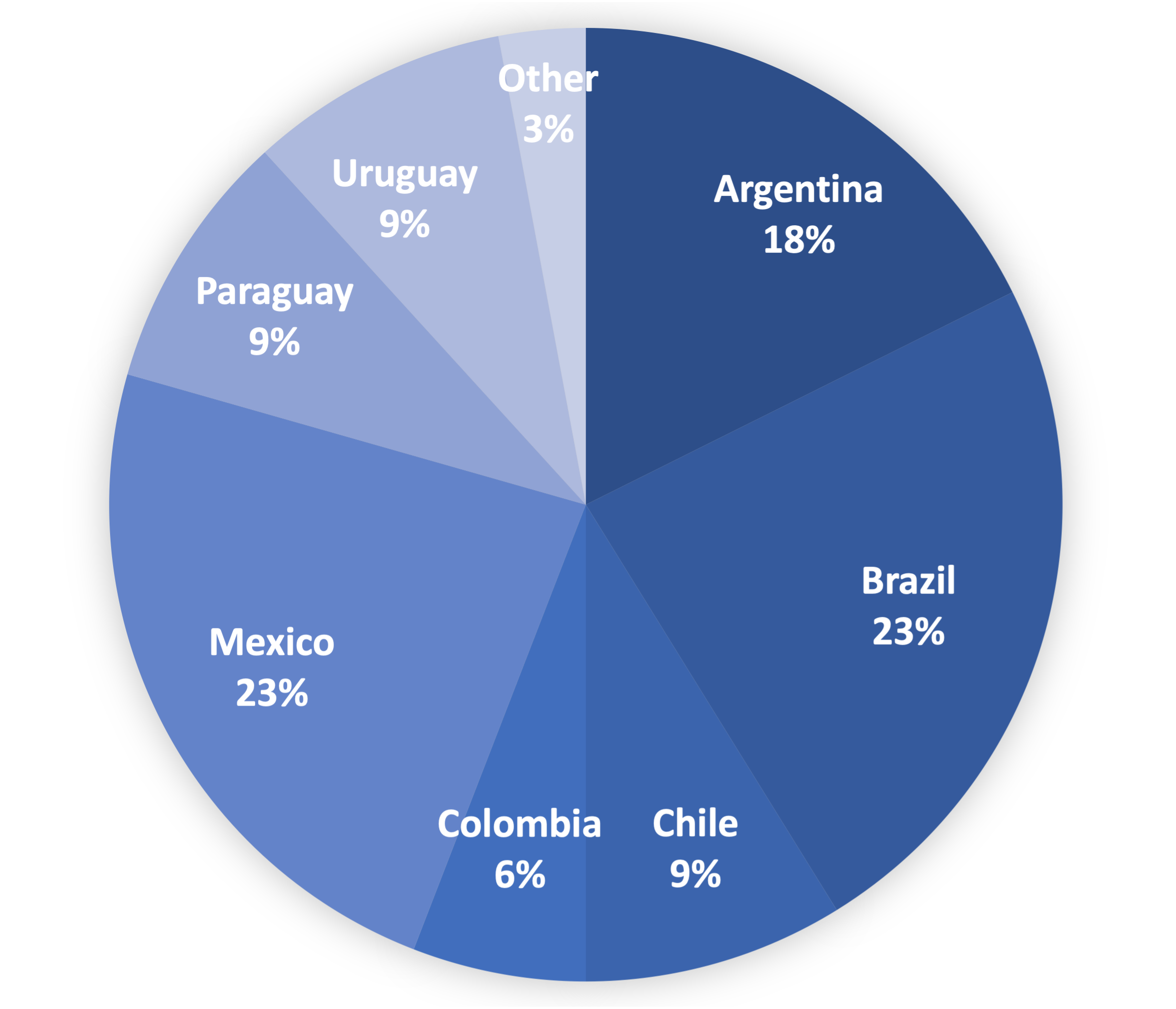}
        \caption{Quantum Computing Research Countries}
        \label{fig:ComputerResearchCountries} 
        
    \end{subfigure}
    \begin{subfigure}[b]{0.48\textwidth} 
        \centering    \includegraphics[width=\textwidth]{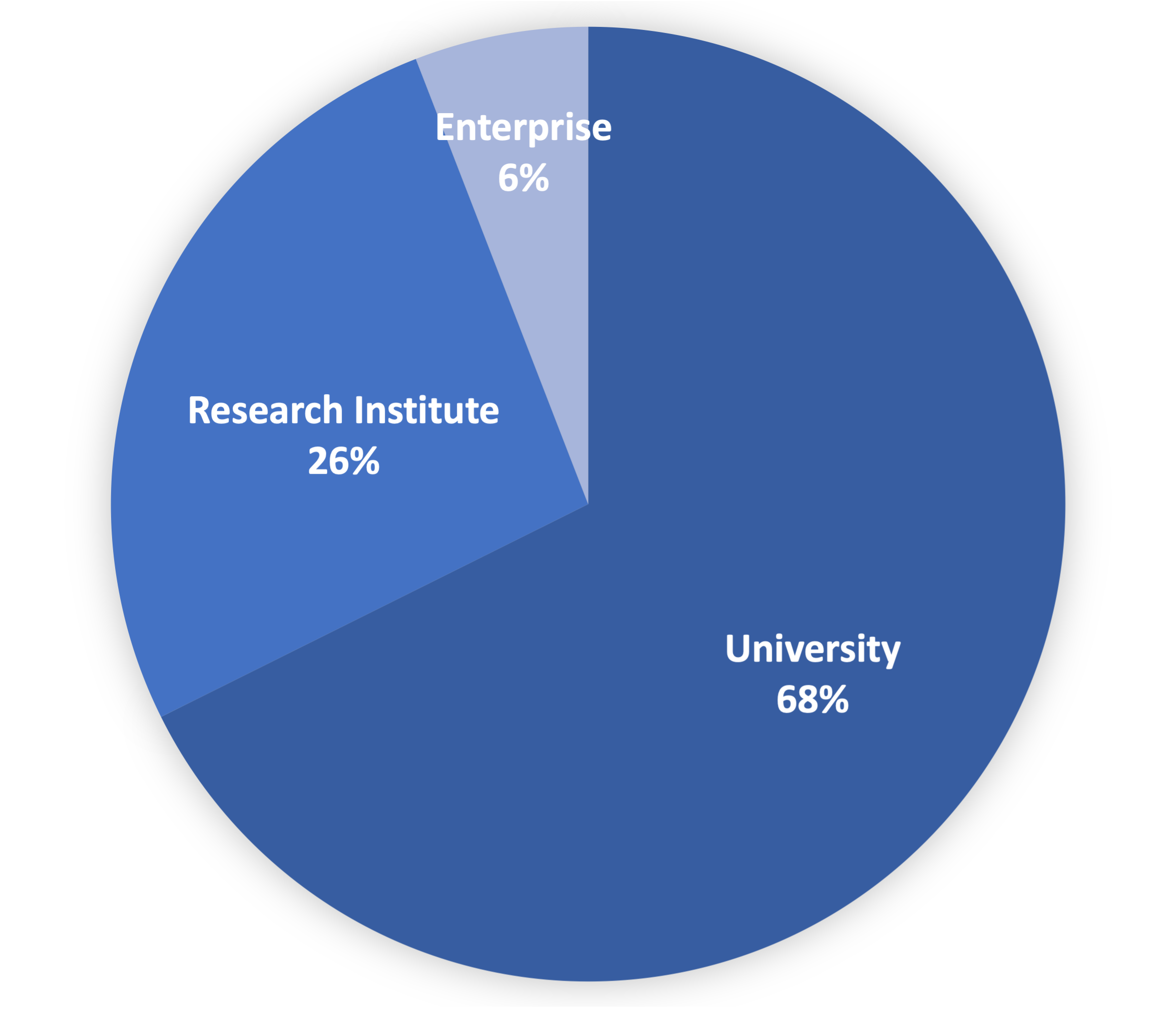}
        \caption{Research Institution}
        \label{fig:ComputerResearchInstitution} 
    \end{subfigure}
    \caption{Quantum Research from Latin American Countries and Institutions} \label{fig:ComputerResearchCountries-Institution}    
\end{figure*}

Figure \ref{fig:ComputerResearchCountries-Institution} provides an overview of the Latin American countries and types of institutions for the survey's participants. As shown in Figure \ref{fig:ComputerResearchCountries}, Brazil, Mexico, and Argentina have higher participation in the survey with 23\%, 23\%, and 18\%, respectively. On the other hand, Figure \ref{fig:ComputerResearchInstitution} shows that 68\% of the survey participants come from university, 26\% from research institutes, while only 6\% come from the private sector. 


\begin{figure*}[h!]
    \centering
    \begin{subfigure}[b]{0.48\textwidth}
        \centering    \includegraphics[width=\textwidth]{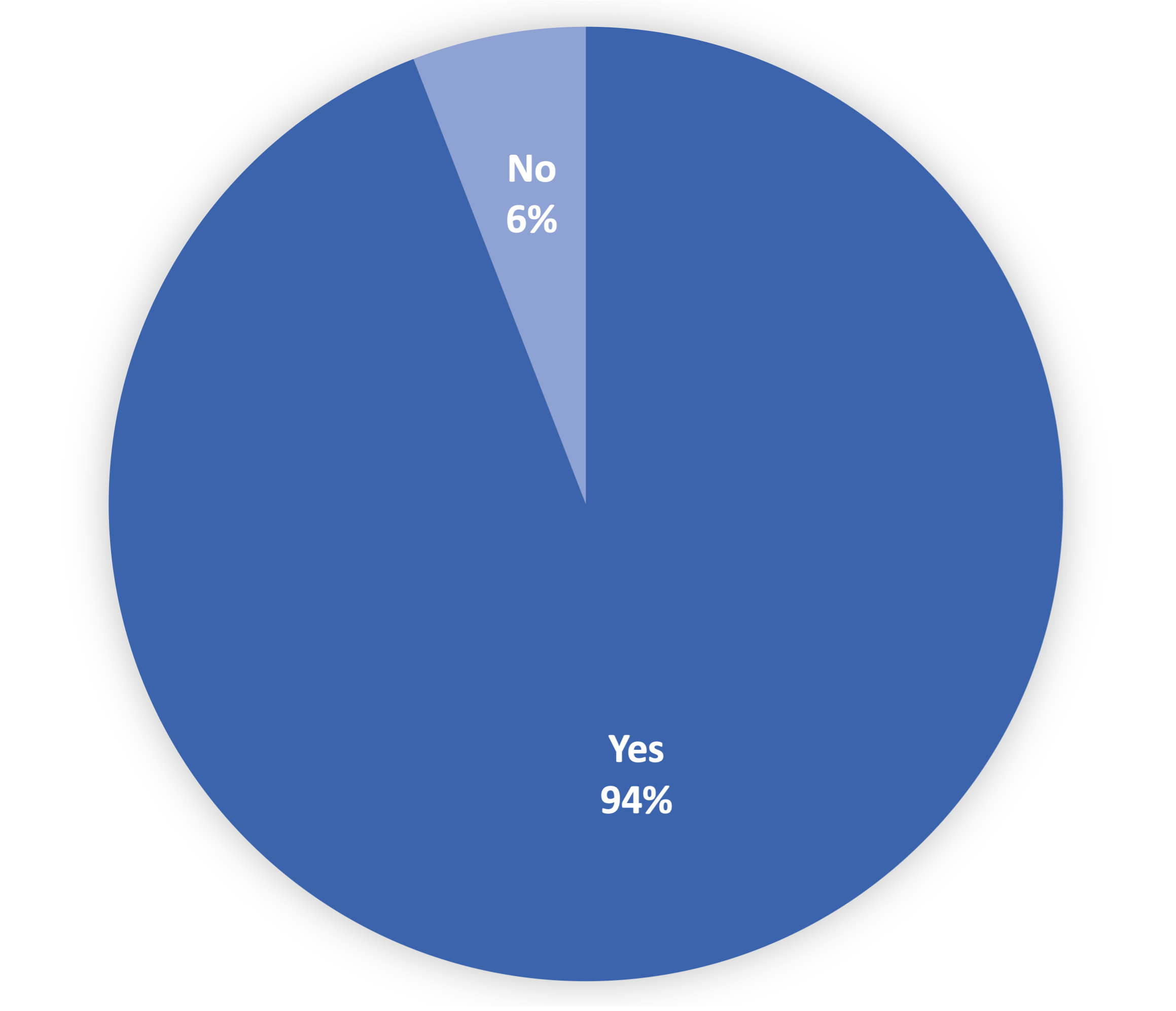}
        \caption{Engaged in Quantum Computing Project}
        \label{fig:Engaged} 
    \end{subfigure}
    \begin{subfigure}[b]{0.48\textwidth} 
        \centering    \includegraphics[width=\textwidth]{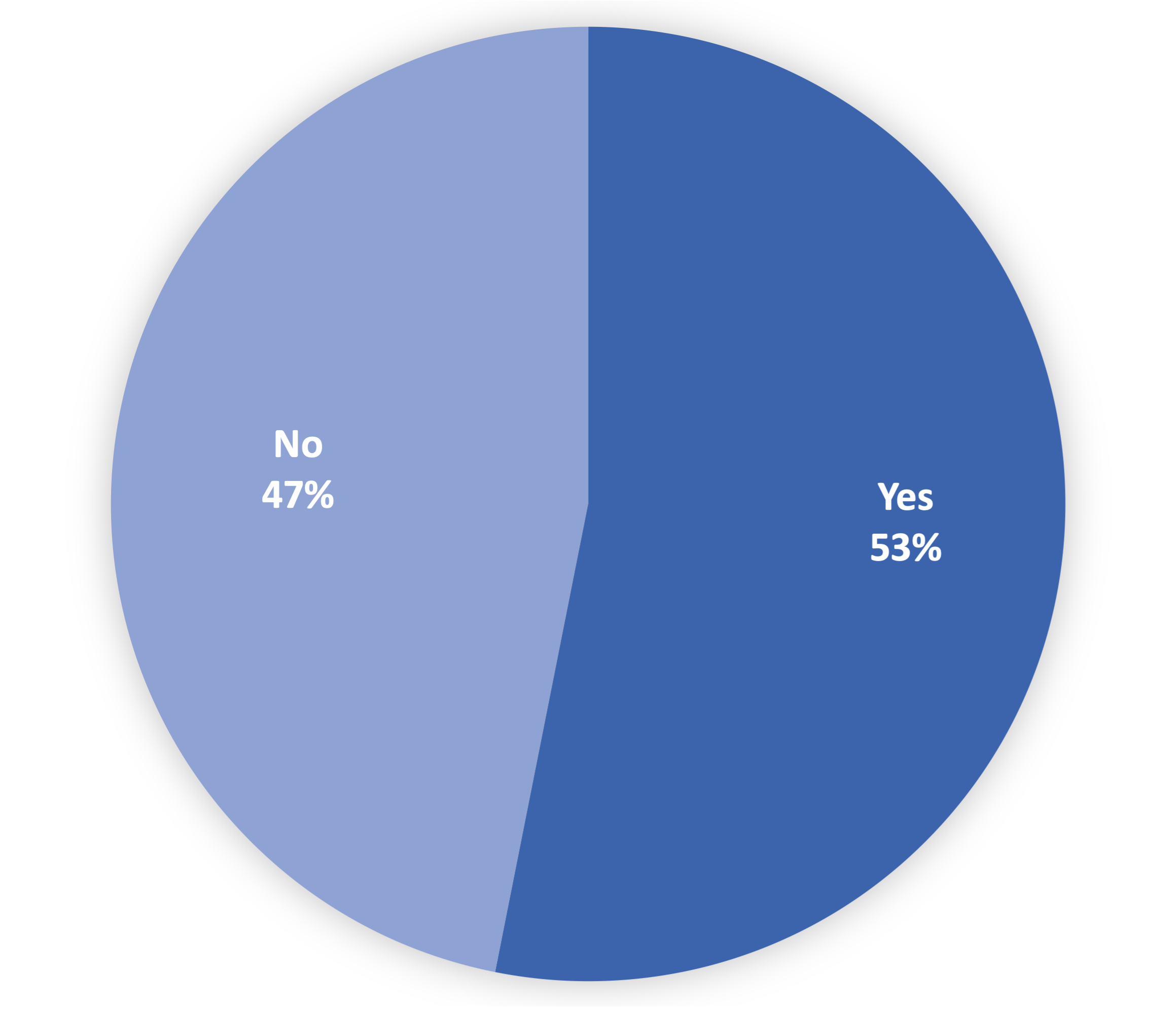}
        \caption{Funded Projects}
        \label{fig:Funded} 
    \end{subfigure}
    \caption{Engaged in Funded Quantum Computing Projects } \label{fig:Engaged-Funded}    
\end{figure*}

On the other hand, we asked respondents whether they are currently involved in quantum computing in any form or topic. The results to that question can be seen in Figure \ref{fig:Engaged}, which shows that 94\% of the respondents are working with aspects related to quantum computing. And of those researchers who are working in the field of quantum computing, 53\% of respondents do so with public funding, as seen in Figure \ref{fig:Funded}. It is worth noting that 35\% of the researchers who indicated that they receive public funding for their research in quantum computing also receive private funding.



\subsection{Research focus}

\begin{figure*}[h!]
    \centering
    \begin{subfigure}[b]{0.49\textwidth} 
        \centering    \includegraphics[width=\textwidth]{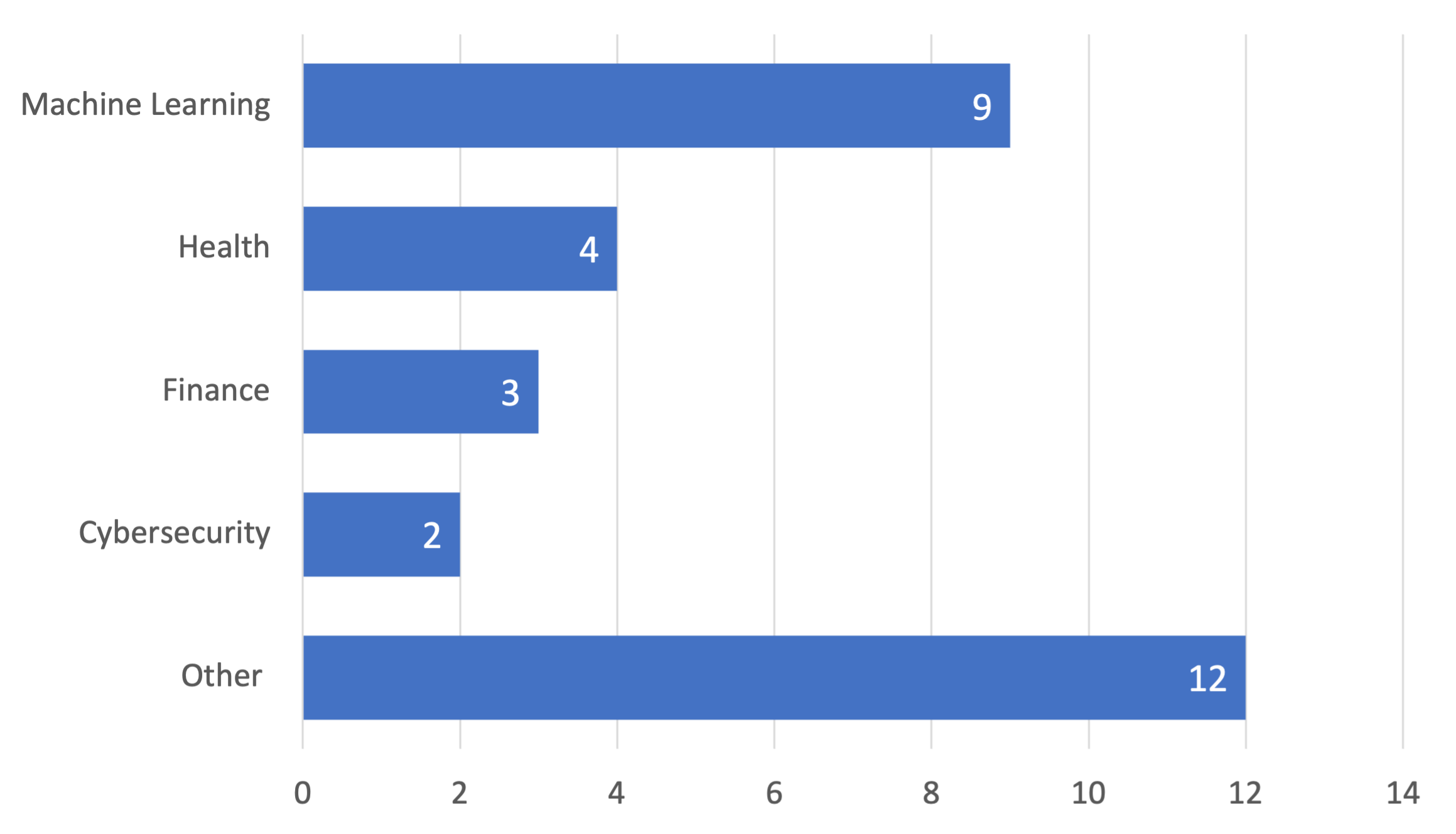}
        \caption{Application domains}
        \label{fig:distribution-by-application-domain.png} 
        
    \end{subfigure}
    \begin{subfigure}[b]{0.49\textwidth} 
        \centering    \includegraphics[width=\textwidth]{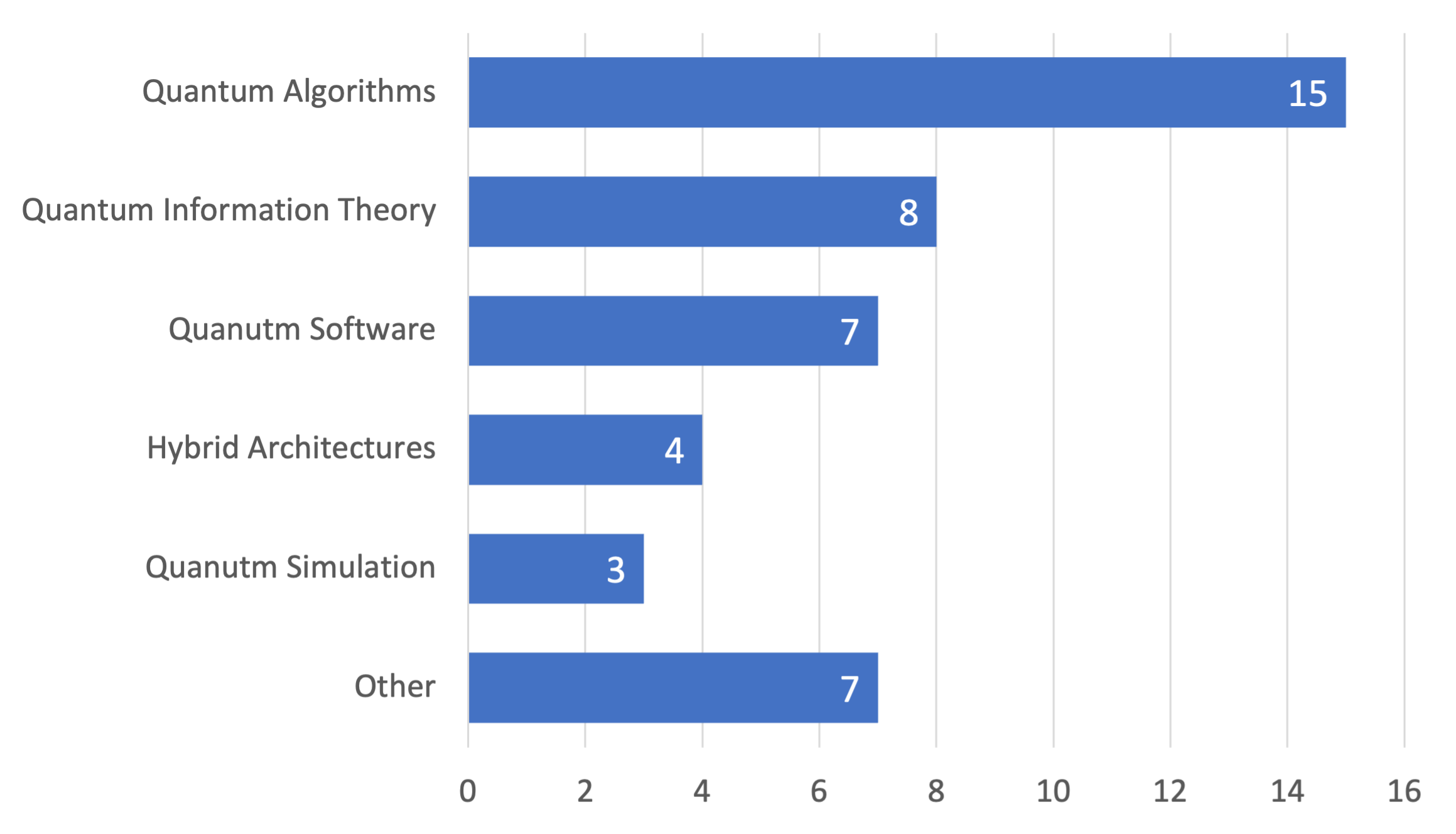}
        \caption{QC topics}
        \label{fig:distribution-by-qc-topic.png}
    \end{subfigure}
    \caption{Focus of respondents already involved in QC research} 
    \label{fig:domains-and-topics}    
\end{figure*}

The researchers involved in QC projects reported that their main fields of application (see Figure \ref{fig:distribution-by-application-domain.png}) are ``Machine learning'' (30\% of the participants), ``Health'', ``Finance'' and ``Cybersecurity'' among others. As depicted in Figure \ref{fig:distribution-by-qc-topic.png}, the most popular quantum computing topics among respondents are ``Quantum Algorithms (New algorithms, Improve existing ones, etc.)'' (34\% of the participants), ``Quantum Information Theory (Entanglement, error correction, etc.)'', and ``Quantum Software (Design, Testing, Modeling, Quality, Development)''.

More than half (22 out of 36) of the participants already involved in QC research recognized working on a topic specific to QSE. When asked specifically about the QSE topic (from a list of predefined options), only three were mentioned: Quantum Software Design, Quantum Software Development, Quantum Software Modelling, and Method. None of the participants mentioned  ``Quantum Software Testing'', ``Quantum Software Requirements'', ``Quantum Software Maintenance'', ``Quantum Software Quality''. These results can be analyzed from different points of view. On one hand, there are researchers working on QSE that do not align themselves with any of the classical software engineering areas of interest. This can imply that QSE covers new areas that are not usually addressed in classical software. On the other hand, some of the classical areas of software engineering are not addressed by any researcher. This can be explained by two different reasons. First, there are areas of classical software engineering that do not need important adaptation for QSE, and therefore pose no interest for researchers. And second, there are also areas of software engineering that are raising significant interest in QSE worlwide but that are not being addressed by Latin American researchers.

\subsection{Collaboration and networking}

Half of the participants report working in collaboration with researchers from other groups or institutions, mainly from their own country and from Europe, and to a lesser extent from Latin America and USA.

\begin{figure}[h!]
    \centering
    \includegraphics[width=0.8\textwidth]{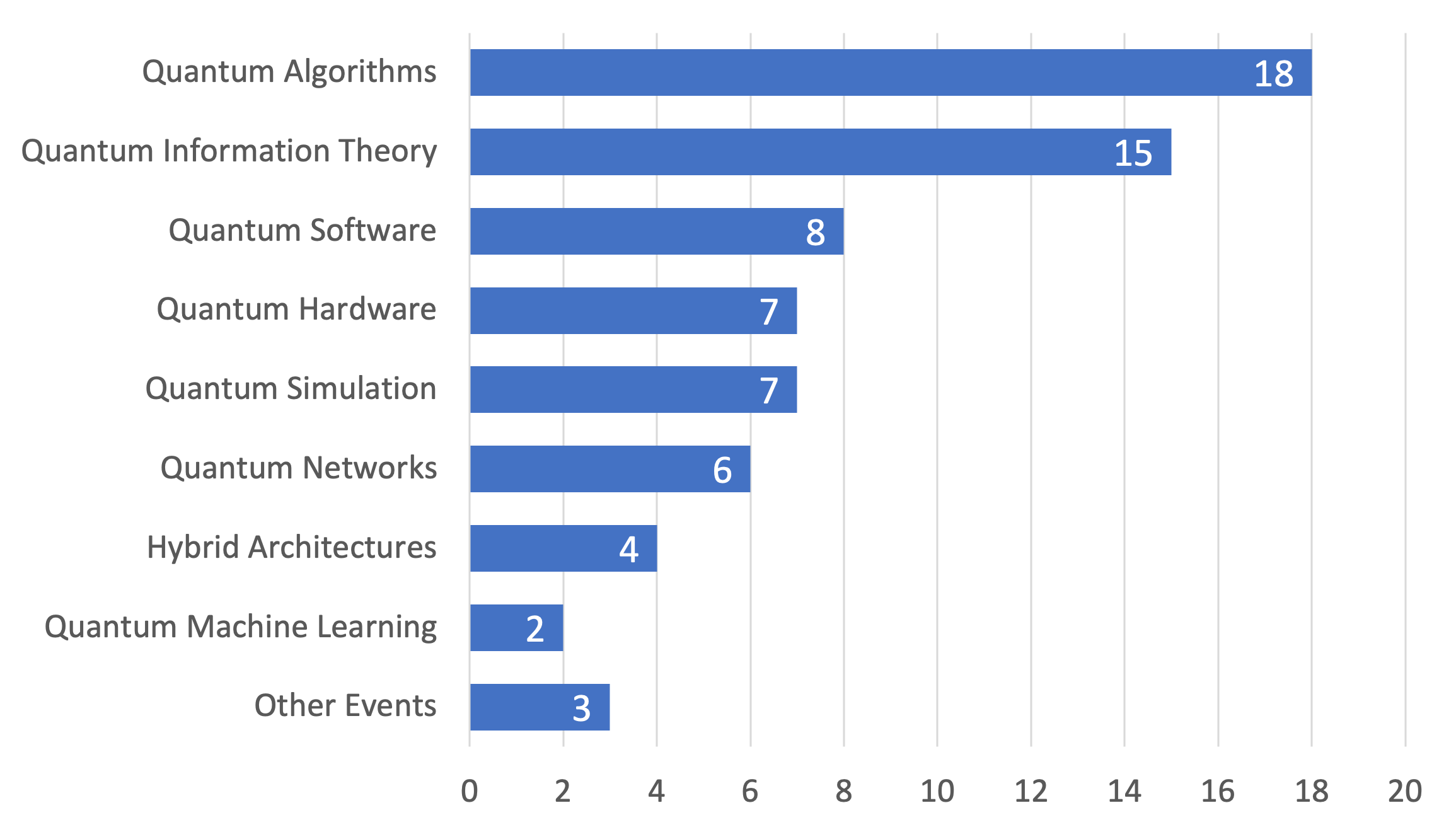}
    \caption{Topics of QC events attended by participants}
    \label{fig:event-topics}
\end{figure}

Two-thirds of the participants report attending Quantum Computing events on varied topics (see Figure \ref{fig:event-topics}). 

\subsection{Relevance awareness and challenges}

To learn about the participants' perception of the relevance of Quantum Software Engineering, we asked them to indicate if they considered it was relevant now or if it will become relevant in the mid-future, in the long term, or only after the consolidation of Quantum Computing. Responses are almost evenly distributed between the four options, with only a small tendency towards the mid and long-term alternatives.  

\begin{figure}[h!]
    \centering
    \includegraphics[width=0.8\textwidth]{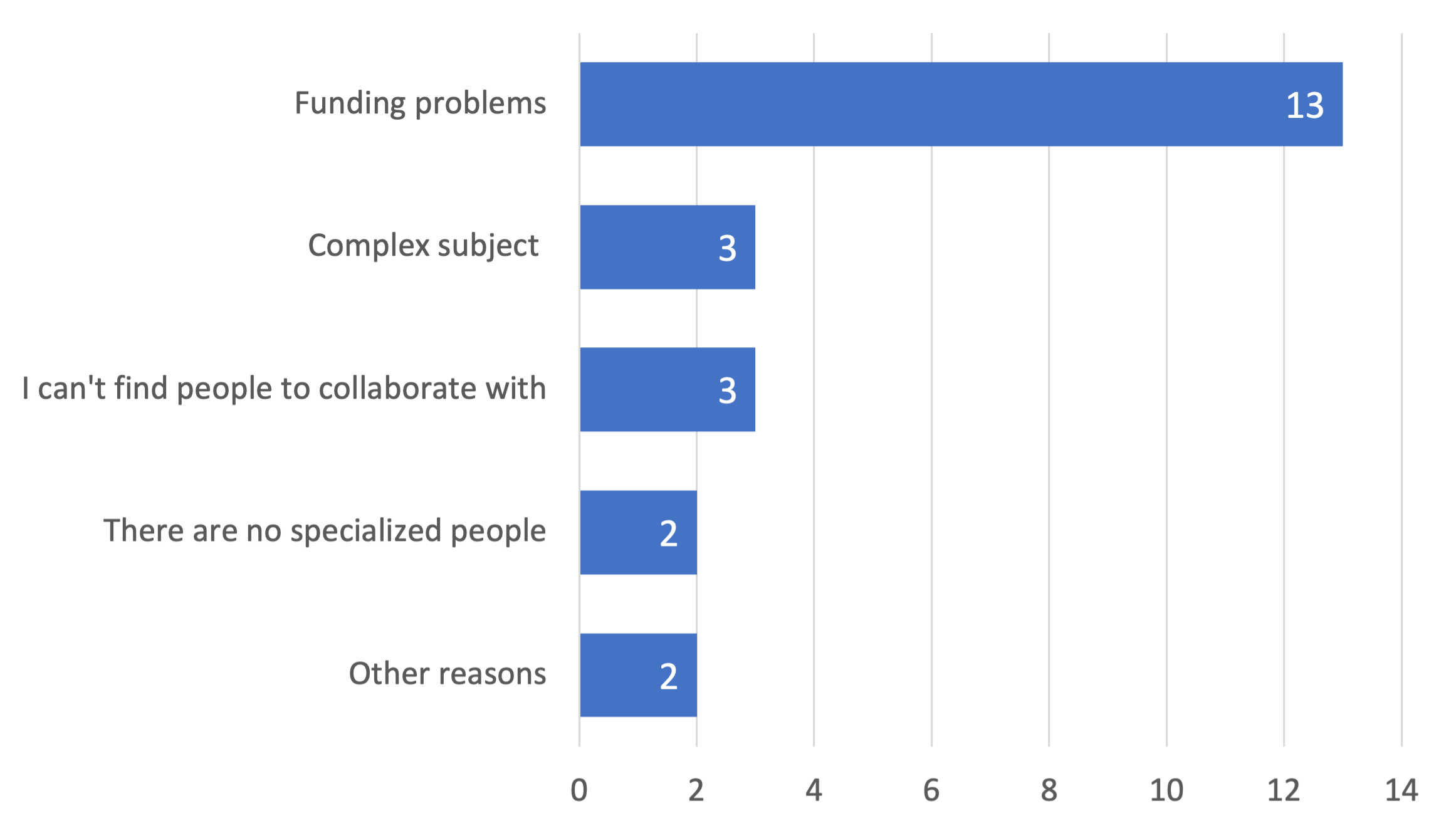}
    \caption{Challenges to get involved in QC research}
    \label{fig:challenges}
\end{figure}

When asked about the challenges they face in getting involved in QSE projects (see Figure \ref{fig:challenges}), participants identified the lack of funding, the complexity of the subject, and the lack of collaboration partners and experts as the main issues.  

\subsection{Discussion}

Once we have analyzed the data obtained from the survey that we have carried out, we would like to point out a possible threat to the validity of the survey itself, which is that the data may be unconsciously biased due to the selection of the respondents. In this sense, we have tried to get the survey to as many researchers in the field of quantum computing as possible, and we believe that the resulting data are close to the reality in which we find ourselves.

To continue, and after performing a knowledge analysis, we can draw several significant conclusions from the point of view of Quantum Software Engineering research in Latin America:  


\begin{itemize}
    \item \textbf{Geographical concentration of research.} There is a direct correlation between the number of survey participants and the authors' affiliation from the systematic review conducted in the previous section. This suggests that the survey reasonably reflects the general opinion of researchers in Latin America. Regarding the concentration of research, two countries (Brazil and Mexico) have the largest number of researchers in this area of knowledge.
    \item \textbf{Crucial role of Universities and Research Institutes.} The overwhelming majority of research participants are affiliated with universities and research institutes. This highlights these institutions' fundamental role in promoting and conducting research activities in the region, especially in cutting-edge research such as in the case.
    \item \textbf{Financial challenges.} It is concerning that only just over half of the participants have funding for their research. This situation may limit the scope and quality of research projects, as well as the ability of researchers to carry out their work effectively.
    \item \textbf{Need for international collaboration.} Although a significant proportion of participants report collaborations with researchers from other countries, collaboration between research groups within Latin America is considerably lower. Fostering and strengthening regional collaboration could be key to improving the quality and impact of research in the region and addressing common challenges.
\end{itemize}

The implications for research communities in Latin America derived from the presented information are diverse and significant. On the one hand, the low percentage of researchers with funding suggests a significant dependency on external funding sources. This may subject researchers to external research agendas and limit their ability to address local issues or develop innovative research lines. On the other hand, the lack of collaboration between research groups within Latin America can restrict the exchange of knowledge and resources among countries in the region. This may result in duplicated efforts, lack of synergies, and reduced capacity to address common challenges effectively. It is crucial to strengthen research infrastructure and institutional support throughout the region to address these implications. This includes increasing funding for research, promoting regional collaboration, and creating conducive environments for developing scientific careers in Latin America. In summary, these implications highlight the importance of addressing the structural challenges facing the research community in Latin America to promote a more equitable, collaborative, and sustainable research environment in the region. This will require a concerted effort from governments, academic institutions, and the scientific community.

Considering the identified implications and current needs in the research community in Latin America, some future directions that could be explored include:

\begin{itemize}
    \item \textbf{Promotion of regional collaboration.} Promoting and facilitating collaboration among research groups within Latin America and researchers from other regions of the world. This can be achieved through exchange programs, research networks, and collaborative projects addressing common challenges.
    \item \textbf{Diversification of funding sources.} Seeking and developing new funding sources for research in Latin America, including public, private, and international funds. This may involve strengthening national funding agencies, promoting public-private partnerships, and participating in international scientific cooperation programs.
    \item \textbf{Strengthening research infrastructure.} Investing in improving and expanding research infrastructure in Latin America, including laboratories, computing centers, libraries, and other resources necessary for conducting high-quality research. This can enhance researchers' capacity to conduct cutting-edge and internationally competitive research.
    \item \textbf{Support for scientific talent development.} Promoting the development of scientific talent in Latin America through training programs, scholarships, and professional development opportunities. This may include creating high-quality graduate programs, supporting academic mobility, and promoting the inclusion of underrepresented groups in science.
\end{itemize}

In summary, these future directions can contribute to strengthening research capacity in Latin America and promoting sustainable scientific and technological development in the region. These initiatives must be developed collaboratively and participatively, involving diverse stakeholders from the scientific, academic, governmental, and business communities.

\section{Conclusions}
\label{sec5}



Quantum computing is postulated as a new computational paradigm capable of addressing several challenges that classical computers cannot. To program quantum computers, quantum software must be developed. As with classical software, quantum software is expected to be large and complex, so Quantum Software Engineering must be defined.

Taking advantage of the special issue focused on Latin America, it is important to know the development being carried out by universities, research institutes, and companies to highlight these advances in QSE. This can serve as a starting point for pioneering initiatives and collaborations between Latin American countries to improve and develop this new area of quantum software engineering, such as the RIPAISC network (\textit{Red Iberoamericana para el Avance de la Ingeniería de Software Cuántico}).

It is essential to establish cooperation between these stakeholders. Therefore, it is necessary to identify who is working on QSE and their research scope within this new field.

To this end, we have carried out a systematic mapping of the literature to identify those papers and researchers coming from Latin American universities or research centers to know their area of research in Quantum Software Engineering. From this literature review, we extracted revealing results, such as the main Latin American countries and researchers in this new research area.

Continuing our study, we surveyed to extract information about the vision and opinion of experts in this field. We sent this survey to the 18 researchers of the RIPAISC network, to 113 researchers who were referred to us by the network researchers themselves, and 39 Latin American researchers from the papers extracted from the systematic mapping study. From this survey, we extracted revealing results such as the fact that most of the researchers in this field come from universities, that more than half of the researchers in this field obtain public funding for their research, or that most of them work in the field of quantum algorithm development; among other results already described.

However, the main conclusion from this comprehensive study is that more collaborative efforts are needed between Latin American researchers.

\backmatter

\bmhead{Supplementary material}

The supplementary material derived from this study can be found on the Zenodo platform (\url{https://doi.org/10.5281/zenodo.10718726}) and is as follows:

\begin{itemize}
    \item From the literature search conducted through a systematic mapping study, we extracted 1316 papers potentially of interest to the study. From this search, we derived the files "\textit{2\_all papers with filters.xlsx}" (title, authors, year of publication, publication forum, DOI, and link) and "\textit{3\_All papers without filters.xlsx}".

    \item Once the inclusion and exclusion criteria were applied, the file "\textit{1\_Selected papers.xlsx}" was extracted. File with the most relevant information of each of the papers selected for our study: title, link, software engineering category, DOI, year of publication, authors' country, and type of contribution.

    \item For this study, we also conducted a complete survey to evaluate the current state of research in the field of quantum software engineering in Latin America and the future claims of researchers in the region. The complete survey can be found in the file "\textit{4\_Survey.pdf}".

    \item The anonymized survey results can be found in the file "\textit{5\_Response to the survey.xlsx}".
\end{itemize}

\bmhead{Acknowledgments}
This work was partially funded by the European Union ``Next GenerationEU /PRTR'' by the Ministry of Science, Innovation and Universities (projects PID2021-1240454OB-C31, TED2021-130913B-I00, and PDC2022-133465-I00). It was also supported by QSERV: Quantum Service Engineering: Development Quality, Testing and Security of Quantum Microservices project funded by the Spanish Ministry of Science and Innovation and ERDF; by the Regional Ministry of Economy, Science and Digital Agenda of the Regional Government of Extremadura (GR21133); and by the European Union under Agreement 101083667 of the Project ``TECH4E -Tech4effiencyEDlH'' regarding the Call: DIGITAL-2021-EDlH-01 supported by the European Commission through the Digital Europe Program. This study was also supported by the QSALUD project (EXP 00135977 / MIG-20201059) in line with the actions of the Center for the Development of Industrial Technology (CDTI). This study was also supported by the ``Ingeniería de Software Cuántico'' project, funded by the Faculty of Informatics of the National University of La Plata.

\bibliography{sn-bibliography}

\end{document}